\documentclass[aps,prl,twocolumn,groupedaddress,showpacs,floatfix,10pt]{revtex4-1}

\usepackage{graphicx}
\usepackage{amssymb}
\usepackage{bm}
\usepackage{verbatim}

\usepackage{color}

\begin{document}


\title{Experimental validation of nonextensive scaling law in confined granular media} 



\author{Ga\"el Combe}
\author{Vincent Richefeu}
\author{Marta Stasiak}
\affiliation{UJF-Grenoble 1, Grenoble-INP, CNRS UMR 5521, 3SR Lab \\ Grenoble F-38041, France}
\email[]{gael.combe@3sr-grenoble.fr}

\author{Allbens P.F. Atman}
\affiliation{Departamento de F\'{\i}sica e Matem\'atica and National Institute of Science and Technology for Complex Systems,
Centro Federal de Educa\c{c}\~ao\ Tecnol\'ogica de Minas Gerais -- CEFET-MG. Av. Amazonas 7675, 30510-000, Belo Horizonte-MG, Brazil.}
\email[]{atman@dppg.cefetmg.br}

\date{\today}

\begin{abstract}
In this letter, we address the relationship between the statistical fluctuations of grain displacements for a full quasistatic plane shear 
experiment, and the corresponding anomalous diffusion exponent, $\alpha$. We experimentally validate a particular case of the so-called 
Tsallis-Bukman scaling law, $\alpha = 2 / (3 - q)$,  where $q$ is obtained by fitting the probability density function (PDF) of the measured 
fluctuations with a $q$-Gaussian distribution, and the diffusion exponent is measured independently during the experiment. 
Applying an original technique, we are able to evince a transition from an anomalous diffusion regime to a Brownian behavior as a function of 
the length of the strain-window used to calculate the displacements of grains in experiments. The outstanding conformity 
of fitting curves to a massive amount of experimental data shows a clear broadening of the fluctuation PDFs  as the  
length of the strain-window decreases, and an increment in the value of the diffusion exponent - anomalous diffusion. Regardless of the size of 
the strain-window considered in the measurements, we show that the Tsallis-Bukman scaling law remains valid, which is the first experimental 
verification of this relationship for a classical system at different diffusion regimes. We also note that the spatial correlations show 
marked similarities to the turbulence in fluids, a promising indication that this type of analysis can be used to explore the origins of 
the macroscopic friction in confined granular materials.
\end{abstract}
\maketitle

Turbulence is one of the most complex, but ubiquitous, phenomena observed in Nature and it is related with the underlying mechanisms 
responsible for the micro-macro upscale causing wide-ranging effects on classical systems, like macroscopic friction in granular 
solids or turbulent flow regime in fluids \cite{Nguyen2014,Wilcox1988,Hou2013,RadjaiRoux2002}.
The presence of multiple scales in time and space is an additional defy to a comprehensive theoretical description, and 
a particular effort is made in the literature to perform experiments and simulations in order to validate the proposed theoretical 
descriptions, particularly Tsallis nonextensive (NE) statistical mechanics \cite{tsallis1988,tsallis2009,combe2013,richefeu2012}. 

A paradigmatic work relating anomalous diffusion and turbulent-like behavior in confined granular media was presented by 
Radjai and Roux \cite{RadjaiRoux2002}, using numerical simulations, and confirmed qualitatively by experiments by Combe and 
collaborators \cite{combe2013,richefeu2012}. Radjai and Roux coined a new expression to 
characterize the analogies between fluctuations of particle velocities in quasistatic granular flows and the velocity fields observed in 
turbulent fluid flow in high Reynolds number regime, the ``granulence''.  Most of the evidences of the granulence are based in 
simulations using discrete element method (DEM) but, unfortunately, one can verify a lack of quantitative 
experimental verification in the last years, limiting the knowledge of the micromechanics of this system based almost exclusively 
on numerical evidences. 

In the present work, we aim exactly to fill this gap by contributing with the experimental validation of the results 
obtained by DEM. Specifically, we seek to examine the findings revealed by Radjai and Roux \cite{RadjaiRoux2002}
in a detailed fashion, extending the previous works \cite{combe2013,richefeu2012} to explore quantitatively the relationship between
the PDF of the velocity fluctuations and the diffusion features of the grains. We follow a detailed theoretical description for the 
anomalous diffusion in the presence of external driving \cite{tsallisbukman1996, plastino1995}. Particularly, a relation between the
$q$-Gaussian value from the PDF of fluctuations and the diffusion exponent was proposed, which is validated experimentally here for the 
first time for a large range of the control parameter, differently of previous works where this relation was tested only for a single point 
\cite{upadhyayaa2001,daniels2004}.

In this work, we aim to advance in the route opened by Radjai and Roux \cite{RadjaiRoux2002} with three basic goals:\\
(\textit{i}) 
\emph{Explore the low inertial number limit.} The inertial number $I$ \cite{RouxCombe2003} measures the ratio between 
inertial and confining forces, from the quasistatic regime (small values) to the dynamic regime (large ones) \cite{GdRMIDI}. We would 
like to check if the granulence features are still observed in a better stablished quasistatic situation, {\textit i.e.} the experimental 
one which involves inertial numbers around four orders of magnitude smaller than the currently reported in simulations  \cite{RadjaiDubois2011}.\\
(\textit{ii}) 
\emph{Point out the origins of the macroscopic friction.} 
We take advantage of the really quasistatic feature of the experimental data to explore the origins of the 
underlying mechanisms of granulence. Here, unlike fluid flow, the rigid particles can not fly
freely since the motion of each particle is hampered by the presence of the other particles, and depends on the motion of its neighbors. 
This makes the straining in part controlled by geometric exclusions at the particle scale, preventing the development of a uniform straining 
in a sustainable way. As shown in \cite{combe2013}, at the limit of large strain-windows, it is possible to observe turbulent-like 
vortexes in the fluctuation field which turn out to be associated with the energy dissipation and macroscopic friction 
\cite{rognon2015,miller2013}.\\
(\textit{iii})  
\emph{Evince the nonextensive nature of the displacement fluctuations.} In order to quantitatively analyze the data, we have used the 
Tsallis NE statistical mechanics approach. In this context, the PDF of displacement fluctuations is not expected to follow the normal 
Gaussian distribution, as in the case of classical Maxwell-Boltzmann distribution in thermodynamics. In granular systems under loading, 
the force chains engaged along the entire system are a clear evidence of long-range interactions \cite{majumdar2005}. These chains 
connect the microscopic contact forces with the global resistance to external forces, as shear for example \cite{Estrada2008}. 
Thus, it is natural to associate the emergence of these force chains at mesoscopic scales with the departure from the classical 
Boltzmann-Gibbs (BG) statistics in these systems.\\

In our experiment, we have foreseen the possibility to quantify the degree of nonextensivity using the $q$-Gaussian fit of the PDF obtained
experimentally \cite{combe2013}. The striking accordance observed on the fitted curves, and the dependence observed of $q$ as a
function of the strain-window used to calculate the fluctuations, according the reasoning presented here, corroborates the application 
of the NE statistical mechanics on these systems, opening an alternative approach to treat these systems quantitatively. Besides, 
by measuring the diffusion of the particles along the complete shear test, at different strain-window, we are able to associate the 
$q$-value measured from the fluctuation PDFs with the diffusion exponent $\alpha$. This is a particular case 
of the Tsallis-Bukman scaling law \cite{tsallisbukman1996},
\begin{equation}
\label{eq:qalpha}
 \alpha = \frac{2}{3-q}\ ,
\end{equation}
which can be obtained from the so-called \emph{porous media equation} \cite{plastino1995}, a generalization of the classical diffusion 
equation where the linear dependence between the variance and time is no longer observed \cite{Poschel2001granular}:
\begin{equation}
 \label{eq:anomdiff}
  \frac{\partial p(x,\,t)}{\partial t} = D_q \frac{\partial^2 \left[p(x,\,t)\right]^{2-q}}{\partial x^2}\ .
 \end{equation}
For a Dirac delta initial condition, the solution reads as
 \begin{equation}
 \label{eq:qgauss}
  p_q(x,\,t) = \frac{1}{\sqrt{\pi A_q } } e_q^{-\frac{x^2 }{ A_q}} \equiv \frac{1}{\sqrt{\pi A_q } } {\left[1 - (1-q)\frac{x^2}{A_q} \right]^{\frac{1}{1-q}}}. 
 \end{equation}
where $e_q(x)$ is called $q$-exponential, and $A_q$ is a constant which depends on $q$ and Gamma-function \cite{plastino1995,tsallis2009}.

\begin{figure}[tb]
\centering
\includegraphics[width=\linewidth]{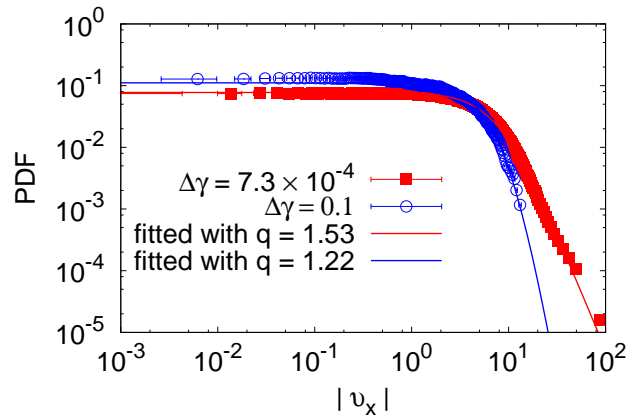}
\caption{Probability density functions of the horizontal components of the fluctuating displacements tracked during two different
increments of shear strain ($\Delta \gamma = 7.3 \times 10^{-4}$ and $\Delta \gamma = 10^{-1}$. The scatters correspond to experimental data, 
and the solid lines correspond to regressions of function $p_q$ (Eq. \ref{eq:qgauss}) that allow for the assessment the q-values.}
\label{fig:PDF}
\end{figure}

Equation \ref{eq:qgauss} is known as the $q$-Gaussian distribution, and was used to fit the PDF of displacement fluctuations obtained 
experimentally. Figure \ref{fig:PDF} shows the results for the PDF of fluctuations and the corresponding fit function at two extremal values of 
$\Delta \gamma$ considered in the image analysis. 

In the case of anomalous diffusion, it is shown that the variance follows a power law with time:
 \begin{equation}
 \label{eq:anomvariance}
  \langle x^2 \rangle \propto  t^{\alpha} \quad \equiv \quad \langle x^2 \rangle \propto t^{\frac{2}{3-q}}\ ,
  \end{equation}
where $\alpha$ is the diffusion exponent equivalently expressed as a function of $q$ by using Eq.~\ref{eq:qalpha}. Note that, 
in Eqs~\ref{eq:anomdiff} to \ref{eq:anomvariance}, $x$ stands for a fluctuation of displacement as we will see below. It is interesting to 
observe two special cases: when $q=1$, the variance is proportional  to time which corresponds to the normal diffusion behavior; when $q=2$, 
the ballistic diffusion limit is reached. At intermediate values, we get large distributions with marked tails. The variance, 
calculated as the time-integral of $p_q$, diverges for $q>5/3$, and converges otherwise. Thus, if several independent convolutions are applied, 
$p_q$ approaches a Gaussian distribution if $q<5/3$, and it approaches a L\'evy distribution for $q>5/3$ \cite{tsallis2009}.

\begin{figure}[tb]
\centering
\includegraphics[width=\linewidth]{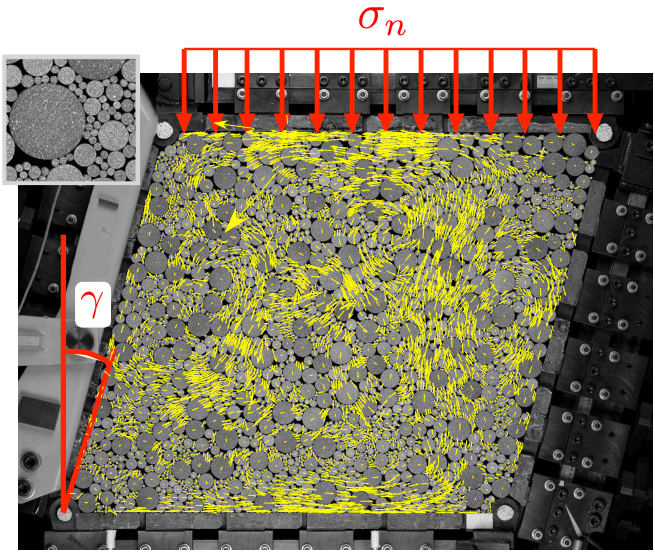}
\caption{The fluctuating part of the rod displacement $\bm{v}(\gamma , \Delta \gamma)$ over-plotted on the corresponding digital image, 
obtained from DIC technique. $\gamma = \Delta \gamma = 0.1$. 
Inset: a detailed view of the speckled rods. $\gamma$ and $\sigma_n$ are the shear angle and the vertical stress imposed, respectively. 
The shear strength is measured all along the shear test.}
\label{fig:system} 
\end{figure}

We performed quasistatic simple shear tests with $1\gamma2\varepsilon $ apparatus which is fully described in \cite{Joer1992,Calvetti1997}. 
The granular packing is made of pilings of cylindrical rods which mimics a 2D granular material enclosed by a rectangular frame, with initial 
dimensions of $0.56\ \textrm{m} \times 0.47\ \textrm{m}$. Then, the vertical sides of this rectangular parallelogram are shortened or 
elongated to apply a constant normal stress in the vertical direction, $\sigma_n = 50$~kPa. These two vertical sides are tilted up to 
$\gamma = 15^\circ$ while the two other sides are kept horizontal with a constant length -- Fig.~\ref{fig:system}. The packing was made 
of $5471$ wooden rollers ($6$~cm long) with ten different diameters ranging from $3$~mm to $30$~mm, approaching to a uniform distribution. 
To ensure a quasi-static transformation of the sample, it is sheared very slowly -- 
the corresponding shear rate $\dot{\gamma}$ is $4.5 \times 10^{-5}$ s$^{-1}$. This ensure a very small inertial number \cite{GdRMIDI} 
($I = 10^{-9}$) when compared to what is applied in DEM simulations ($10^{-3}$ to $10^{-5}$ in the best cases)
\cite{RouxCombe2010,RadjaiDubois2011}. During the test, kinematics of 
grains are measured by means of Digital Image Correlation (DIC) \cite{he1984two,chu1985ap} from 80~MPixels digital images of the sample 
where rollers look like disks, Fig. \ref{fig:system}. A specific \emph{DIC} computer program was developed to track rollers here assumed 
as rigid bodies \cite{richefeu2012} which allowed a sub-pixel kinematics measurement to track grains with an error of $\pm 0.05$ pixels 
\cite{combe2013tracker}.

At the macroscopic level (sample scale), the stress-strain curve measured during the shear test exhibited hardening up to 
$\gamma \approx 0.06$ followed by softening until the end of the test (curve shown in \cite{richefeu2012} as well as several other 
mechanical properties like peak stress ratio, macroscopic friction angle \emph{etc}.). 
Pictures were shoot every $\delta t = 5$~s throughout  the test, corresponding to a shear strain increment 
$\Delta \gamma \equiv \delta t\, \dot{\gamma} \approx 2.4 \times 10^{-4}$ between each shot.
To assess the displacement fluctuations, we consider two displacements of each particle during a shear increment  $\Delta \gamma$. The first 
is the actual displacement $\delta \bm{r}(\gamma,\, \Delta \gamma)$ from $\gamma$ to $\gamma + \Delta \gamma$. The second displacement, 
$\delta \bm{r}^\star(\gamma,\, \Delta \gamma) $, is fictitious and corresponds to an affine motion resulting from an homogeneous straining at 
$\gamma$ and during the shear-increment  $\Delta \gamma$. It is assessed from the motion of the four rigid sides of the apparatus 
$1\gamma2\varepsilon$. With these definitions, the fluctuating part of the displacement is the difference between the actual and affine 
displacements. Thus, the normalized displacement fluctuation $\bm{v}(\gamma,\, \Delta \gamma)$ is defined by:
\begin{equation}
\bm{v}(\gamma,\, \Delta \gamma) = \frac{\left [ \delta \bm{r}(\gamma,\, \Delta \gamma) - \delta\bm{r}^\star(\gamma, \, \Delta \gamma) \right ] / d}{\Delta \gamma} \; ,
\end{equation}
where $d$ is the mean diameter of the rollers. One may notice that the normalized fluctuations can be interpreted as a local strain 
(grain scale -- numerator) compared to the global strain (sample scale -- denominator).

%

%
A displacement fluctuation field $\bm{v}$ is plotted in Fig.~\ref{fig:system} for a given shear increment. We notice an organization in 
structures like vortexes that reminds the ones of turbulence in fluids. These structures found their origins in the rearrangement mechanism 
of the grains, since that the elements interfere with each other in their affine movement. This is, in other words, 
the deviation from the affine field due to steric exclusion forming patterns observed with discrete element modeling 
\cite{Kuhn1999,RouxCombe2002,RouxCombe2003} and more rarely in experiments \cite{miller2013}.
Their dynamics depend both on $\gamma$ and $\Delta \gamma$, evolving gently under shear when 
$\Delta \gamma$ is large ($>0.04$) and very rapidly for small values ($\Delta \gamma \simeq 2.4\cdot 10^{-4}$). The characteristic lengths 
depend strongly on $\Delta \gamma$, with vortexes of a few tenths of grain mean diameter for large values of $\Delta \gamma$, and, to the 
contrary, for small values of $\Delta \gamma$ these structures are not well defined, and long range correlations are observed 
\cite{richefeu2012}.

The PDFs of the horizontal component magnitude of normalized displacement fluctuations are shown in 
Fig.~\ref{fig:PDF} for two different increments of shear strain: $\Delta \gamma = 7.3 \cdot 10^{-4}$ and $\Delta \gamma = 10^{-1}$. 
We observe a broadening of the PDF from a nearly Gaussian distribution (for large $\Delta \gamma$) to a wider distribution 
(for small $\Delta \gamma$). 


\begin{figure}[t]
\centering
\includegraphics[width=\linewidth]{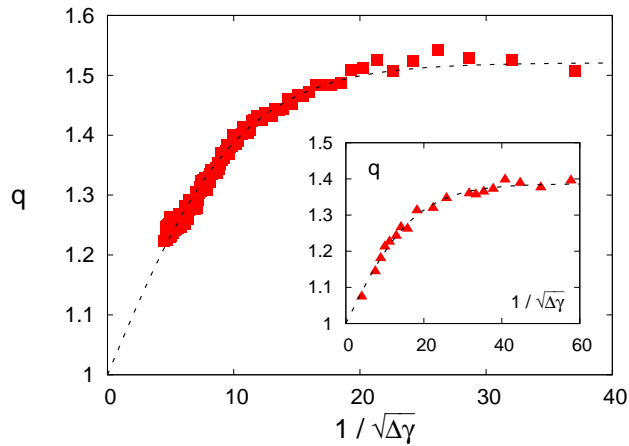}
\caption{ 
Evolution of the measured $q$-value as a function of squared inverse of the strain increment for the experiments and simulations. 
The dashed line corresponds to a regression with the function  $q(1/\Delta \gamma) = 1 + a tanh(b / \Delta \gamma)$, with
$a = 0.521$, $b = 0.096$. Inset: the same plot for data from a DEM simulation that highlights the limit $q = 1$ when 
$\Delta \gamma \rightarrow \infty$. The fitted parameters for simulations were $a = 0.387$, $b = 0.057$.}
\label{fig:qevolution} 
\end{figure}

The dependence of the $q-$exponent with the strain-window used to calculate the fluctuations is shown in Fig.~\ref{fig:qevolution}, for 
experimental and simulation data. Two remarkable features can be observed in this plot:
first, in the limit of large strain-window, when the abscissa goes to zero, $q \rightarrow 1$, indicating the limit when normal diffusion and 
the BG statistics are satisfied. Note that it is possible to test larger values of $\Delta \gamma$ in DEM simulations which confirm 
the limit $q \rightarrow 1$ (data shown in the inset of Fig. \ref{fig:qevolution}).
This is exactly what we expect for this limit, once that the particles typically experience 
several collisions and rearrangements, approaching to the molecular chaos hypothesis. 

In the other limit, for vanishing strain-window, the $q$-value attains a plateau, with $q \sim 3/2$. This observation can be interpreted 
as a sign of the long range correlations imposed by the force chains at this short time scale.
Once the value measured for $q$ in this limit is lower than $5/3$, one can expect that for large strain-windows a Gaussian distribution 
would be recovered, since it correspond to successive independent convolutions of $q$-Gaussian distributions.

These features were observed both in experiments and simulations, no matter the differences among the systems (periodic boundaries in horizontal 
direction in simulations, different number of particles and inertial numbers \emph{etc}), proving the robustness of the result.

Analyzing the results as a whole, we can sketch a phenomenological scenario to explain the observations: in the limit of large $ \Delta \gamma$, we 
observe a tendency to agree with the BG statistics, with $q \rightarrow 1$. This limit corresponding to the transition from meso- to macroscopic 
scales, and we observe the formation of vortexes in the spatial distribution of fluctuations, as evinced by Fig. \ref{fig:system}. These 
vortexes, with few grains diameters in size, interact each other to dissipate the excess of energy due to external loading, in analogy with 
the role of the vortexes in turbulent flow \cite{RadjaiRoux2002}. The nature of the interactions of these structures is purely stochastic, which acts as a 
precursor for the macroscopic friction. The broadening of the displacements fluctuations distribution is usually attributed 
to the energy cascade from larger to lower scales, that is, from large vortexes to the small ones \cite{RadjaiRoux2002}. 

\begin{figure}[tb]
\centering
\includegraphics[width=\linewidth]{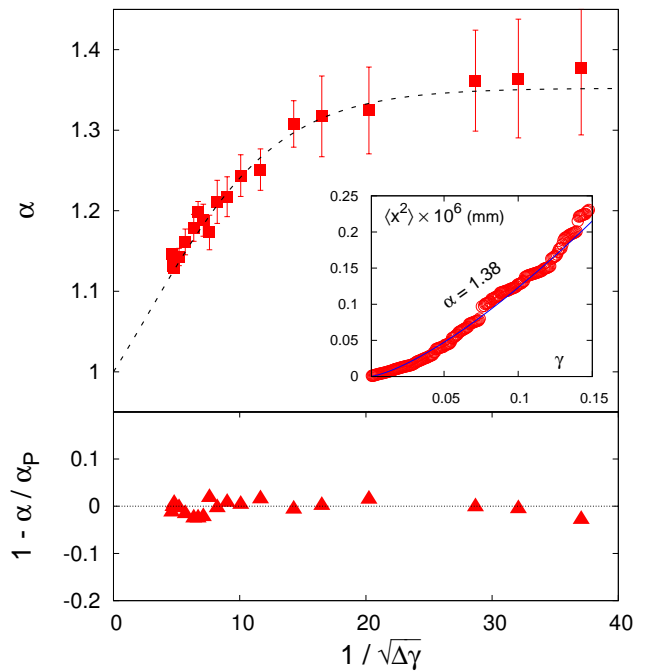}
\caption{
Verification of the Tsallis-Bukman scaling law for different regimes of diffusion. (top) Evolution
 of the measured diffusion exponent $\alpha$ as a function of $1/ \Delta \gamma$ the dashed line
is a direct application of the scaling law from the fit of the values shown in Fig. \ref{eq:qgauss},
$\alpha_P(1/ \Delta \gamma) = 2/[3 - q(1/ \Delta \gamma)]$, where $P$ stands for predicted. (Inset) a selected curve diffusion where $x$ is the displacement fluctuations; 
it allows the assessment of the diffusion exponent $\alpha$. (Bottom) Measure of the deviation of the data
 relative to the scaling law prediction, as a function of $1/ \Delta \gamma$, showing a agreement on the order of $\pm 2\%$.}
\label{fig:qalpha} 
\end{figure}

On the other hand, we have $q \simeq 3/2$ for vanishing strain-windows limit, $\Delta \gamma \rightarrow 0$. This result indicates the 
presence of long-range interactions and anomalous diffusion. Considering the absence of spatial structures on the fluctuation field, it is clear that  this limit is 
dominated by the force chain dynamics. Force chains can span all along the system, but are very fragile, implying  short  life times. 
The displacement of grains belonging to a force chain are strong correlated spatially, but this correlation is not verified for temporal scales. 

Thus, we can conclude that the window used to measure the PDF particle displacement fluctuations in the system plays a crucial role in 
the statistics that will be obtained. Basically, it is possible to explore the micro-macro transition on the PDF distributions, from a correlated 
regime dominated by the force chains to a frictional stochastic one, dominated by spatial vortex interactions. This conclusion has a striking 
implication for any analysis concerning the measuring of displacement fluctuations, since it unveils how the observation procedure 
can alter the conclusions even in a relatively simple diffusion experiment.
 
To quantify the diffusion of the grains along the complete shear test we basically computed the 
average displacement of each grain as a function of time (shear increment $\gamma$), but with different sampling frequencies determined 
by the strain-window $\Delta \gamma$. Following the reasoning presented above, and Eq.~\ref{eq:anomvariance}, we should expect two extreme 
regimes for the diffusion, analogously to the observed for the $q$-value: an anomalous diffusion regime 
with $\alpha \sim 4/3$ for vanishing strain-window, and an asymptotic regime with $\alpha \rightarrow 1$ for large shear increments. 
This is indeed what we can observe in Fig. \ref{fig:qalpha}, where we have verified the Tsallis-Bukman scaling law (Eq. \ref{eq:qalpha}).
It is important to stress that the dashed line in Fig. \ref{fig:qalpha} is \textbf{not} a direct fit, but rather the curve obtained in Fig.
\ref{fig:qevolution} using the Tsallis-Bukman scaling law. 
To our knowledge, is the first time that this relation is verified for different regimes of diffusion.
This striking result reinforces the use of the Tsallis NE statistical mechanics to describe strong correlated systems, as in the case of 
confined granular material under shearing.

\begin{acknowledgments}
We are indebted to Constantino Tsallis for the fruitful discussions, suggestions and kind reading of the manuscript. We thank Philippe
Claudin for the kind reading of the manuscript and suggestions. 
We are grateful to Jean-Beno\^it Toni for his valuable work to upgrade the electronic part of $1\gamma2\varepsilon$ apparatus. A special thanks 
to Fran\c{c}ois Bonnel without whom we would nott have the chance to shoot with the Phase One IQ180 camera (80 MPixels).
APFA thanks the  Brazilian funding agencies FAPEMIG, CNPq and CAPES, 
and CEFET-MG for financial support. The Laboratoire 3SR is part of the LabEx Tec 21 (Investissements d'Avenir - grant agreement n$^\textrm{o}$ ANR-11-LABX-0030)
\end{acknowledgments}

\bibliography{cra}
\end{document}